\documentclass[12pt]{article}
\usepackage{latexsym}
\usepackage{amsmath,amsbsy}

\newcommand{\dd}{\mbox{\rm d}}
\newcommand{\wg}{\wedge}
\newcommand{\gam}{\gamma}
\newcommand{\Gam}{\Gamma}

\newcommand{\ddg}{\ddagger}
\newcommand{\tl}{\tilde}

\newcommand{\bgam}{\boldsymbol{\gamma}}

\newcommand{\DD}{\mbox{\rm D}}

\newcommand{\nnn}{\noindent}
\newcommand{\oo}{\over}
\newcommand{\p}{\partial}

\newcommand{\be}{\begin{equation}}
\newcommand{\bear}{\begin{eqnarray}}
\newcommand{\ear}{\end{eqnarray}}
\newcommand{\ee}{\end{equation}}
\newcommand{\lbl}{\label}

\newcommand{\ci}{\cite}

\newcommand{\vs}{\vspace}

\begin{document}

\

\baselineskip .6cm 

\vs{27mm}

\begin{center}

{\LARGE \bf Spin Gauge Theory of Gravity

\vs{2mm}

in Clifford Space}\footnote{Talk presented at the
QG05 conference, 12-16 September 2005, Cala Gonone, Italy}

\vs{3mm}

Matej Pav\v si\v c

Jo\v zef Stefan Institute, Jamova 39,
1000 Ljubljana, Slovenia

e-mail: matej.pavsic@ijs.si

\vs{6mm}

{\bf Abstract}

\end{center}

\vs{2mm}

A theory in which 16-dimensional curved Clifford space ($C$-space) provides
a realization of Kaluza-Klein theory is investigated. No extra
dimensions of spacetime are needed: ``extra dimensions" are in $C$-space.
We explore the spin gauge theory in $C$-space and show that the generalized
spin connection contains the usual 4-dimensional gravity and
Yang-Mills fields of the U(1)$\times$SU(2)$\times$SU(3) gauge
group. The representation space for the latter group is provided
by 16-component generalized spinors composed of four usual 4-component
spinors, defined geometrically as the members of four independent
minimal left ideals of Clifford algebra.

\vs{8mm}

\section{Introduction}

Spacetime geometry can be elegantly described by means of geometric
calculus in which basis vectors are generators of Clifford
algebra\,\ci{Hestenes}.
This enables a description of geometric objects of different
grades, i.e., multivectors or $r$-vectors associated with oriented
$r$-dimensional surfaces. In previous works it has been shown
\ci{PavsicArena, AuriliaFuzzy} that multivectors can sample extended
objects, e.g., closed branes, and that in this respect they generalize
the notion of center of mass. Instead of describing an extended
object by an infinite number of degrees of freedom, which is one extreme,
or only by the center of mass coordinates, which is another extreme,
one can describe it by a finite number of multivector coordinates
which take account of object's extension and orientation. Such
description works for macroscopic objects as well, if we assume
that they are composed of branes, which is a reasonable assumption
within the framework of a theory based on strings and branes. Then
the multivectors associated with the constituent branes sum together
to give an effective multivector describing the macroscopic object
\ci{PavsicArena}.

The basis multivectors span a $2^n$-dimensional space, called
{\it Clifford 
space} or $C$-space, $n$ being the dimension of the underlying space
that we start from. We will assume that the starting space is 4-dimensional
spacetime. A point in $C$-space is described by a set of
multivector coordinates $(\sigma,~ x^\mu,~x^{\mu \nu},...)$ which
altogether with the corresponding basis elements $({\bf 1}, \gam_\mu,
\gam_{\mu \nu},...)$ form a Clifford aggregate or {\it polyvector} $X$.
It is well known that the elements of the right or left minimal ideals
of Clifford algebra can be used to represent spinors. Therefore,
a coordinate polyvector $X$ automatically contains not only
bosonic, but also spinor coordinates. In refs.\,\ci{PavsicParis,PavsicSaasFee}
it was proposed to formulate string theory in terms of polyvectors, and
thus avoid usage of a higher dimensional spacetime. Spacetime can be
4-dimensional, whilst the extra degrees of freedom (``extra dimensions")
necessary for consistency of string theory are in $C$-space.

There is a fascinating possibility of a generalization from {\it flat}
$C$-space, serving as an arena for physics, to {\it curved} $C$-space which
is itself a part of the play 
\ci{PavsicParis,PavsicKaluza1,PavsicKaluza2,CastroUnific}.
Since a dynamical (curved) Clifford space has 16 dimensions, it provides
a realization of Kaluza-Klein idea. We do not need to assume that spacetime
has more than four dimensions. The ``extra dimensions" are in Clifford
space, and they are all physical, because they are associated with the degrees
of freedom related to extended objects (see also \ci{PavsicArena}).
So we do not need to ``compactify" or in whatever way to hide them.

\section{Clifford space}

In a series of preceding works \ci{Pezzaglia}--\ci{CastroPavsicReview}
it has been proposed to construct the extended relativity theory
in $C$-space by a natural generalization of the notion of spacetime interval:
\be
    \dd S^2 = \dd \sigma^2 + \dd x_\mu \dd x^\mu + \dd x_{\mu_1 \mu_2}
    \dd x^{\mu_1 \mu_2} + ... + \dd x_{\mu_1 ...\mu_n} \dd x^{\mu_1 ... \mu_n}
\lbl{2.1}
\ee
where $\mu_1 < \mu_2 < ... < \mu_n$. The Clifford valued polyvector\footnote
{If we do not restrict indices according to $\mu_1 < \mu_2 < ... $, then
the factors $1/2!,~1/3!,...,$respectively, have to be included in front
of every term in eq.\,(\ref{2.2}).}
\be
  X= x^M \gam_M = \sigma {\bf 1} + x^\mu \gam_\mu + x^{\mu_1 \mu_2}
  \gam_{\mu_1 \mu_2} + ... + x^{\mu_1 ...\mu_n} \gam_{\mu_1 ...\mu_n}
\lbl{2.2}
\ee
denotes the position of a point in a manifold, called Clifford space
or $C$-space. The series of terms in eq.\,(\ref{2.2}) terminates at
finite grade depending on the dimension $n$ of spacetime that we start
from. A Clifford algebra ${\cal C} \ell_{r,q}$ with $r+q=n$ has $2^n$ basis
elements. Here we keep $n$ and the signature arbitrary, but later
we will take $n=4$, and signature $+ - - -$.

In {\it flat} $C$-space one can choose a basis so that the relation
with wedge product
\be
    \gam_M = \gam_{\mu_1} \wg \gam_{\mu_2} \wg ... \wg \gam_{\mu_r} \
    \equiv \frac{1}{r!} [\gam_{\mu_1},\gam_{\mu2},...,\gam_{\mu_r}]
\lbl{2.3}
\ee
$\gam_\mu,~ \mu=1,2,...,n$, being the generators of Clifford algebra,
holds globally for all points $X$ of $C$-space.

The interval (\ref{2.1}) can be written as the {\it scalar product}
\be
    \dd S^2 = \dd X^\ddg * \dd X = \dd x^M \dd x^N \, G_{MN} =
    \dd x^M \dd x_M
\lbl{2.4}
\ee
of $\dd X = \dd x^M \gam_M$
with its reverse $\dd X^\ddg$. {\it Reversion} is an operation which
reverses the order of vectors, e.g., $(\gam_1 \gam_2 \gam_3)^\ddg =
\gam_3 \gam_2 \gam_1$. The scalar product between two polyvectors $A$
and $B$ is defined according to
     $A * B = \langle AB \rangle_0$,
where $\langle ~\rangle_0$ denotes the zero grade part.

The {\it metric}, entering the quadratic form (\ref{2.4}) is given by
\be
    G_{MN} = \gam_M^\ddg * \gam_N
\lbl{2.7}
\ee
If the underlying spacetime $V_n$ has signature $(+ - - - - ....)$, then
the Clifford space has signature $(+ + + ... - - - ..)$ with equal number
of plus and minus signs. This has some potentially far reaching
consequences, discussed in refs.\,\ci{PavsicParis,PavsicSaasFee}.

As in the ordinary theory of relativity we generalize flat (Minkowski)
spacetime to curved spacetime, so we now generalize flat $C$-space
to {\it curved $C$-space}. We can use analogous concepts and techniques.

A set of $2^n$ linearly independent polyvector fields on a region
${\cal R}$ of $C$-space will be called a {\it frame field}. Of
particular interest are:
\begin{description}
    \item{~(i)} {\it Coordinate frame field} $\{\gam_M\}$. Basis elements
    $\gam_\mu$, $M= 1,2,...,2^n$ depend on position $X$ in $C$-space.
    The relation (\ref{2.3}) with wedge product can hold only
    locally at a
    chosen point $X$, but in general it cannot be preserved globally
    at all points $X \in {\cal R}$ of {\it curved} $C$-space.
    The scalar product of two basis elements determines the metric tensor
    of the frame field $\{\gam_M \}$
    according to eq.\,(\ref{2.7}).
    
    \item{(ii)} {\it Local flat frame field} $\{\gam_A\}$.
    Basis elements $\gam_A,~A=1,2,...,2^n$ also depend on $X$, but
    at every point $X$ they determine {\it flat} metric
\be 
    \gam_A^\ddg * \gam_B = \eta_{AB}
\lbl{2.8} \ee
\end{description}

The relation between the two sets of basis elements is given in term of
the $C$-space vielbein:
\be
    \gam_M = {e_M}^A \gam_A
\lbl{2.9}
\ee
All quantities in eq.\,({2.9}) depend on position $X$ in $C$-space.
Reciprocal basis elements $\gam^M$ and $\gam^A$ are defined according to
$(\gam^M)^\ddg * \gam_N = {\delta^M}_N$ and $(\gam^A)^\ddg * \gam_B =
{\delta^A}_B$.

Let us now define a differential operator $\p_{\gam_M} \equiv \p_M$,
which will be called {\it derivative}, whose action depends on the
quantity it operates on\footnote{This operator is the $C$-space
analogue of the derivative $\p_{\gam_\mu} \equiv \p_\mu$ which operates
in an $n$-dimensional curved space $V_n$, and was defined by Hestenes
\ci{Hestenes} (who used a different symbols, namely $\Box_\mu$).}:
\begin{itemize}

\item $\p_\mu$ maps scalars $\phi$ into scalars
\be \p_M \phi = \frac{\p \phi}{\p x^M} \lbl{2.10} \ee
Then $\p_M$ is just the ordinary partial derivative.

\item $\p_M$ maps Clifford numbers into Clifford numbers. In particular
it maps a coordinate basis Clifford number $\gam_N$ into another Clifford
number:
\be
    \p_M \gam_N = \Gam_{MN}^J \gam_J
\lbl{2.11}
\ee
The above relation defines the {\it coefficients of connection} for
the coordinate frame field $\{\gam_M\}$.
\end{itemize}

An analogous relation we have for the local flat frame field:
\be
      \p_M \gam_A = - {{\Omega_A}^B}_M \, \gam_B
\lbl{2.12}
\ee
where ${{\Omega_A}^B}_M$ are the coefficients of connection for the
local flat frame field $\{ \gam_A \}$.

Expanding an arbitrary polyvector field according to
    $A = A^M \gam_M = A^M {e^A}_M \, \gam_A$
and using eqs.\,(\ref{2.11}),(\ref{2.12}) we have
\be
    \p_N {e^C}_M - \Gam_{NM}^J \, {e^C}_J - {e^A}_M \, {{{\Omega}_A}^C}_N
    = 0
\lbl{2.14}
\ee
which is analogous to the well known relation in an ordinary curved
spacetime, and ${{{\Omega}_A}^C}_N$ extends the notion of spin connection
${{\omega_a}^c}_\mu$.

From (\ref{2.14}) we obtain
\be
    \p_M {e^C}_N - \p_N {e^C}_M + {e^A}_M {{\Omega_A}^C}_N 
    - {e^A}_N {{\Omega_A}^C}_M = {T_{MN}}^J {e^C}_J
\lbl{2.15}
\ee
where ${T_{MN}}^J  = \Gam_{MN}^J - \Gam_{NM}^J$ is the $C$-space torsion.

Taking the commutators of derivatives we have
\be
[\p_M,\p_N] \gam_J = {R_{MNJ}}^K \gam_K
\qquad {\rm and} \qquad
   [\p_M,\p_N] \gam_A = {R_{MNA}}^B \gam_B
\lbl{2.17}
\ee
where
\be
 {R_{MNJ}}^K= \p_M \Gam_{NJ}^K - \p_N \Gam_{MJ}^K + 
 \Gam_{NJ}^R \Gam_{MR}^K -  \Gam_{MJ}^R \Gam_{NR}^K
\lbl{2.18}
\ee
\be
    {R_{MNA}}^B= -(\p_M {{\Omega_A}^B}_N - \p_N {{\Omega_A}^B}_M
    + {{\Omega_A}^C}_N {{\Omega_C}^B}_M - {{\Omega_A}^C}_M {{\Omega_C}^B}_N)
\lbl{2.19}
\ee
are the coefficients of curvature for the frame field $\{\gam_M\}$ and
$\{\gam_A\}$, respectively.

\section{The generalized Dirac equation in $C$-space}

We will leave aside further discussion of a classical general relativity
in $C$-space 
(see \ci{PavsicBook,CastroPavsicHigher,PavsicKaluza1,PavsicKaluza2}) and
go directly to quantum theory. We will assume that wave functions are
polyvector valued fields.

Let $\Phi (X)$ be a polyvector valued
field over coordinates polyvector
field $X= x^M \gam_M$:
\be
    \Phi = \phi^A \gam_A
\lbl{3.1}
\ee
where $\gam_A,~A=1,2,...,16$, is a local (flat) basis of $C$-space
(see eq.(\ref{2.2})) and $\phi^A$  the projections (components) of
$\Phi$ onto the basis $\lbrace \gam_A \rbrace$. We will suppose that in
general $\phi^A$ are complex-valued scalar quantities.
We will assume
\ci{PavsicBook,PavsicKaluza2} that the imaginary unit $i$ is the
bivector of {\it phase space}, and that it commutes with all elements
of ${\cal C} \ell_{1,3}$, since it is not an element of the latter algebra.

Instead of the basis $\lbrace  \gam_A \rbrace$ one can consider another
basis, which is obtained after multiplying $\gam_A$ by 4 independent
primitive idempotents \ci{Teitler}
$P_i = {1\oo 4} ({\bf 1} +a_i \gam_A)({\bf 1}+b_i \gam_B) \; , 
    \quad i=1,2,3,4$. Here $a_i,~b_i$ are complex numbers chosen
so that $P_i^2 = P_i$. For explicit and systematic
construction see \ci{Teitler}.

By means of $P_i$ we can form minimal ideals of Clifford algebra. A
basis of left (right) minimal ideal ${\cal I}_i^L$ (${\cal I}_i^R$)
is obtained by taking 
$P_i$ and multiplying it from the left (right) with all 16 elements
$\gam_A$ of the algebra:
   \be \gam_A P_i \in {\cal I}_i^L \; , \qquad P_i \gam_A \in {\cal I}_i^R
\lbl{3.3}
\ee   
For a fixed $i$ there are
16 elements $\gam_A P_i \in {\cal I}_i^L$ (or $P_i \gam_A \in 
{\cal I}_i^R$,  but only 4 amongst
them are different, the remaining elements are just
repetition---apart from constant factors---of those 4 different elements.

Let us denote those different element that form a basis of the i-th
left ideal by symbol $\xi_{\alpha i},~\alpha = 1,2,3,4$. Altogether,
for $i=1,2,3,4$, there are 16 different basis elements $\xi_{\alpha i}$.
Every Clifford number can be expanded either in terms of $\gam_A$, or in
terms of $\xi_{\tl A} \equiv \xi_{\alpha i} = (\xi_{\alpha 1},
\xi_{\alpha 2}, \xi_{\alpha 3},\xi_{\alpha 4})$ according to
\be
   \Psi = \psi^{\tilde A} \xi_{\tilde A} = \psi^{\alpha 1} \xi_{\alpha 1}
  + \psi^{\alpha 2} \xi_{\alpha 2} + \psi^{\alpha 3} \xi_{\alpha 3} +
\psi^{\alpha 4} \xi_{\alpha 4}
\lbl{3.4}
\ee
Eq.(\ref{3.4}) represents a sum of four independent
4-component {\it spinors}, each in a different left ideal ${\cal I}_i^L$.
We have introduced a single spinor index ${\tl A}$ which runs over all
16 basis elements $\xi_{\tl A}$ that span 4 independent left minimal
ideals of ${\cal C} \ell_{1,3}$. The set $\{ \xi_{\tl A} \}$ of 16 linearly
independent fields $\xi_{\tl A} (X)$ will be called {\it generalized
spinor frame field}. An explicit relation between the two basis is
given in refs.\,\ci{Teitler, PavsicKaluza1,PavsicKaluza2}.

In refs.\,\ci{PavsicCliff,PavsicBook} it was
proposed that the
polyvector valued wave function satisfies
the Dirac equation in $C$-space:
\be
     \p \Psi \equiv \gam^M \p_M \Psi = 0
\lbl{3.5}
\ee
The derivative $\p_M$ is the same derivative introduced in
eqs. (\ref{2.10})--(\ref{2.12}). Now it acts on the object
$\Psi$ which, according to eq.\,(\ref{3.4}), is expanded in terms of
the 16 basis elements
$\xi_{\tilde A}$  which, in turn,
can be written as a superposition
of basis elements $\gam_A$ of ${\cal C} \ell_{1,3}$. The action of
$\p_M$ on $\gam_A$ is given in (\ref{2.12}). An analogous expression
holds if $\p_M$ operates on
the spinor basis elements $\xi_{\tilde A}$:
\be
    \p_M \xi_{\tilde A} = {{\Gam_M}^{\tilde B}}_{\tl A} \xi_{\tl B}
\lbl{3.6}
\ee
where ${{\Gam_M}^{\tilde B}}_{\tl A}$ are components of the generalized
{\it spin connection}, i.e., the components of the connection of curved
$C$-space for the generalized spinor frame field $\{\xi_{\tl A}\}$.
Thus eq.\,(\ref{3.5}) can be written as
\be
    \gam^M \p_M (\psi^{\tl A} \xi_{\tl A}) = \gam^M (\p_M \psi^{\tl A}
  + {{\Gam_M}^{\tl A}}_{\tl B} \psi^{\tl B}) \xi_{\tl A} 
    \equiv \gam^M (\DD_M \psi^{\tl A})\, \xi_{\tl A} = 0
\lbl{3.7}
\ee
We see that in the geometric form of the generalized Dirac equation (\ref{3.5})
spin connection is automatically present through the operation
of the derivative $\p_M$ on a {\it polyvector} $\Psi$.

An action which leads to eq.\,(\ref{3.5}) is (for a more detailed treatment
see ref.\,\ci{PavsicKaluza2}):
\be
    I[\Psi, \Psi^\ddg] = \int \dd^{2^n} x \, \sqrt{|G|} \, 
    i \Psi^\ddg \p \Psi = \int \dd^{2^n} x\, \sqrt{|G|} \, i
   {\psi^*}^{\tl B} \xi_{\tl B}^\ddg \gam^M \xi_{\tl A} \DD_M \psi^{\tl A}
\lbl{3.8}
\ee
where $\dd^{2^n} x \, \sqrt{|G|}$ is the invariant volume element of the
$2^n$-dimensional $C$-space, $G\equiv {\rm det} \,\, G_{MN}$ being the
determinant of the $C$-space metric.

A generic transformation in the tangent $C$-space $T_X C$ which maps
a polyvector $\Psi$ into another polyvector $\Psi'$ is given by
($\Sigma_{AB}$ being generators defined in
ref.\,\ci{PavsicKaluza1,PavsicKaluza2})
\be
   \Psi' = R \Psi S
\lbl{3.9}
\ee
where  $R = {\rm e}^{{1\oo 4} \Sigma_{AB} \alpha^{AB}}$ and 
$S= {\rm e}^{{1\oo 4} \Sigma_{AB} \beta^{AB}}$,
with
$\alpha^{AB}$ and $\beta^{AB}$ being parameters of the transformation.
It can be shown \ci{PavsicKaluza1,PavsicKaluza2} that in matrix form
the transformation (\ref{3.9}) reads
\be
    \psi'^{\tl A} = {U^{\tl A}}_{\tl B} \psi^{\tl B}
\quad {\rm or} \quad
    \psi' = {\bf U} \psi
~~, \qquad {\bf U} = {\bf R} \otimes {\bf S}^{\rm T}
\lbl{3.10}
\ee
where ${\bf R}$ and ${\bf S}$ are
$4 \times 4$ matrices representing the Clifford numbers $R$ and $S$.
 We see
that the matrix ${\bf U}$ is the direct product of ${\bf R}$ and
the transpose  ${\bf S}^{\rm T}$
of ${\bf S}$, and it belongs, in general, to the group $GL(4,C) \times GL(4,C)$,
which is then subjected to further restrictions resulting from the requirement
that $\Psi^{\ddg} * \Psi$ be invariant, which implies $R^\ddg R = 1$ and
$S^\ddg S = 1$. Then it can be shown\,\ci{PavsicKaluza2}
that the scalar part of the action (\ref{3.8}) is invariant under
local transformations (\ref{3.9}),(\ref{3.10}). Besides ordinary Lorentz
transformations the latter group contains the ``internal" transformations.
The group is large enough to contain the subgroup 
U(1)$\times$SU(2)$\times$SU(3). Whether this indeed provides a description
of the standard model remains to be fully investigated. But there is further
evidence in favor of the above
hypothesis in the fact that a polyvector field $\Psi = 
\psi^{\tl A} \xi_{\tl A}$
has 16 complex components. Altogether it has 32 real components. This
number matches, for one generation, the number of independent states for
spin, weak isospin and color, i.e., $(e,\nu),(u,d)_{b,r,g}$,
together with the corresponding antiparticle
states, in the standard model.

Under a transformation (\ref{3.10}) the covariant derivative and
the spin connection transform, respectively, according to
\be
 {\DD}'_{M} \psi'^{\tl A} = {U^{\tl A}}_{\tl B} \, \DD_M \psi^{\tl B}
\;, \quad {\rm i.e.}, \quad
\DD'_M \psi' = {\bf U}\, \DD_M \psi
\lbl{3.12}
\ee  
\be
    {\Gamma_{M {\tl A}}}^{\tl B} = {U_{\tl D}}^{\tl B}
     {U^{\tl C}}_{\tl A} {\Gamma'_{M {\tl C}}}^{\tl D} +
      \p_M {U^{\tl D}}_{\tl A}\, {U_{\tl D}}^{\tl B}
\; ,
      \quad {\rm i.e.}, \quad
      {\bf \Gamma}_M = {\bf U}\, {\bf \Gamma'_M}\, {\bf U}^{-1} + 
    {\bf U}\, \p_M \, {\bf U}^{-1}
\lbl{3.11}
\ee
where $\DD'_{M}\psi'^{\tl A} = \p'_{M} \psi'^{\tl A} +
      {{\Gam'_{M}}^{\tl A}}_{\tl B} \psi'^{\tl B}$ and
      $\DD_{M}\psi^{\tl A} = \p_M \psi^{\tl A} +
      {{\Gam_{M}}^{\tl A}}_{\tl B} \psi^{\tl B}$.
We see that ${\bf \Gam}_M$ transforms as a non abelian gauge field.
We have thus demonstrated that the generally covariant Dirac equation in
16-dimensional curved $C$-space contains the coupling of spinor
fields $\psi^{\tl A}$ with non abelian gauge fields
${{\Gam_M}^{\tl A}}_{\tl B}$ which
altogether form components of connection in the generalized spinor basis.

\section{The gauge field potentials and gauge field strengths}

We can express the spin connection in terms of
the generators $\Sigma_{AB}= {f_{AB}}^C \gam_C \,$:
\be \Gam_M = {1\oo 4} \,{\Omega^{AB}}_N\, \Sigma_{AB} = {A_M}^A \gam_A \;
, \qquad {A_M}^A = {1\oo 4} \,{\Omega^{CD}}_N \, {f_{CD}}^A
\lbl{4.1}
\ee
The matrices representing the Clifford numbers $\gam^M$ and $\Gam_M$
can be calculated according to\,\ci{PavsicKaluza2}
\be
   \langle {\xi^{\tl A}}^\ddg \gam^M \xi_{\tl B} \rangle_S  = 
   {{(\gam^M)}^{\tl A}}_{\tl B} \equiv {\bgam}^M \quad  {\rm and} \quad
   \langle {\xi^{\tl A}}^\ddg \Gam_M \xi_{\tl B} \rangle_S  = 
   {{\Gam_M}^{\tl A}}_{\tl B} \equiv {\bf \Gam}_M
\lbl{4.1a}
\ee   
The $C$-space Dirac equation, written in matrix form, can be split
according to
\be
   \bgam^{M} (\p_{M} +
{\bf \Gam}_{M}) =
   [ \bgam^\mu (\p_\mu + {\bf \Gam}_\mu ) + \bgam^{\bar M} (\p_{\bar M} +
   {\bf \Gam}_{\bar M})] \psi = 0
\lbl{4.2}
\ee
where $M=(\mu,{\bar M})$, $\mu = 0,1,2,3;~ {\bar M} = 5,6,...,16$.

From eq.\,({\ref{4.1}) we read that the gauge fields $\Gam_M$ contain:
(i) {\it The spin connection} of the 4-dimensional gravity $\Gam_\mu^{(4)}=
{1\oo 8} \,{\Omega^{ab}}_\mu [\gam_a,\gam_b]$.
(ii) {\it The Yang-Mills fields} ${A_\mu}^{\bar A} \gam_{\bar A}$, where
we have split the local index according to $A=(a,{\bar
A})$. For the scalar part $A_\mu^{\underline {\bf o}} 
\gam_{\underline {\bf o}}
\equiv A_\mu {\bf 1}$ we have just the U(1) gauge field.
(iii) {\it The antisymmetric potentials}
${A_M^{\underline {\bf o}}} \equiv A_M = (A_\mu,\,A_{\mu \nu},\,
A_{\mu \nu \rho},\,
A_{\mu \nu \rho \sigma})$, if we take indices $A={\underline {\bf o}}$
 (scalar) and $M= \mu,\mu \nu,\mu \nu \rho,\mu \nu \rho \sigma$.
(iv) {\it Non abelian generalization} of the antisymmetric potentials
$A_{\mu \nu ...}^A$\,.

The $C$-space spin connection thus contains all physically interesting
fields, including the antisymmetric gauge field potentials which
occur in string and brane theories.

Using (\ref{3.6}) we can calculate the curvature according to
     $~~~~~~~[\p_M,\p_N] \xi_{\tl A}= {{R_{MN}}^{\tl B}}_{\tl A}\, 
     \xi_{\tl B}$,
     
\nnn where ${{{R_{MN}}^{\tl B}}}_{\tl A} = 
   \p_M {{\Gam_N}^{\tl B}}_{\tl A} -
    \p_N {{\Gam_M}^{\tl B}}_{\tl A} + 
    {{\Gam_M}^{\tl B}}_{\tl C} {{\Gam_N}^{\tl C}}_{\tl A} -
    {{\Gam_N}^{\tl B}}_{\tl C} {{\Gam_M}^{\tl C}}_{\tl A}$,
    or, in matrix
notation, ${\bf R}_{MN}= \p_M \boldsymbol{\Gamma}_N - 
\p_N \boldsymbol{\Gamma}_M
   + [\boldsymbol{\Gam}_M,\boldsymbol{\Gam}_N]$.
Using (\ref{4.1}), and renaming $R$ into $F$ we have
\be
   {F_{MN}}^A = \p_M {A_N}^A - \p_N {A_M}^A + {A_M}^B {A_N}^C {C_{BC}}^A
\lbl{4.6}
\ee
where ${C_{BC}}^A$ are the structure constants of ${\cal C} \ell_{1,3}$
satisfying $[\gam_A,\gam_B] = {C_{AB}}^C \gam_C$.

\section{Conclusion}

In current approaches to quantum gravity the starting point is often in
assuming that at short distances there exists an underlying structure,
based, e.g., on string and branes, or spin networks and spin foams. It is
then expected that the smooth spacetime manifold of classical general relativity
emerges as a sufficeintly good
approximation at large distances. However, it is feasible to assume that
what will emerge is in fact not just spacetime, but spacetime with certain
additional structure. The approach discussed in this contribution suggests
that the long distance approximation to a more fundamental structure is
Clifford space. This is the space of the degrees of freedom that describe
extended objects. Since Clifford space is a higher dimensional space, it
can serve for a realization of Kaluza-Klein theory, and since all its
dimensions are physically observable to us, there is no need for
compactification of
extra dimensions.

\end{document}